\documentclass[square,numbers]{article}

\usepackage[preprint]{neurips_2023}
\usepackage{anyfontsize}
\usepackage{amsmath}
\usepackage[utf8]{inputenc} 
\usepackage[T1]{fontenc}    
\usepackage[hidelinks]{hyperref}       
\usepackage{url}            
\usepackage{booktabs}       
\usepackage{booktabs}       
\usepackage{array}          
\usepackage{longtable,lscape}      
\usepackage{multirow}
\usepackage{pdflscape}      
\usepackage{amsfonts}       
\usepackage{nicefrac}       
\usepackage{microtype}      
\usepackage{pifont}         
\usepackage{xcolor}         
\usepackage[pdftex]{graphicx, adjustbox} 
\usepackage{natbib}
\usepackage{tikz}
\usepackage{colortbl}
\usepackage{placeins} 
\usepackage{notoccite}

\definecolor{darkgreen}{rgb}{0, 0.5, 0}
\begin{document}

\title{Causality analysis of electricity market liberalization on electricity price using novel Machine Learning methods 
}

\author{
  Orr Shahar\textsuperscript{1} \\
  \And
  Stefan Lessmann\textsuperscript{1,2} \\
  \And
  Daniel Traian Pele\textsuperscript{2,3}\\
}

\maketitle

\vspace{-2ex}
\begin{center}
\textsuperscript{1}Humboldt University of Berlin, School of Business and Economics, Berlin, Germany \\
\textsuperscript{2}Bucharest University of Economic Studies, Bucharest, Romania\\ 
\textsuperscript{3}Institute for Economic Forecasting, Romanian Academy, Bucharest, Romania\\
\end{center}
\vspace{2ex}

\begin{abstract}
\fontsize{14}{14}\selectfont Relationships between the energy and the finance markets are increasingly important. Understanding these relationships is vital for policymakers and other stakeholders as the world faces challenges such as satisfying humanity's increasing need for energy and the effects of climate change. In this paper, we investigate the causal effect of electricity market liberalization on the electricity price in the US. 
By performing this analysis, we aim to provide new insights into the ongoing debate about the benefits of electricity market liberalization. We introduce Causal Machine Learning as a new approach for interventions in the energy-finance field. The development of machine learning in recent years opened the door for a new branch of machine learning models for causality impact, with the ability to extract complex patterns and relationships from the data. We discuss the advantages of causal ML methods and compare the performance of ML-based models to shed light on the applicability of causal ML frameworks to energy policy intervention cases. We find that the DeepProbCP framework outperforms the other frameworks examined. In addition, we find that liberalization of, and individual players' entry to, the electricity market resulted in a 7\% decrease in price in the short term. The code is available on https://github.com/Orri11/causality\_analysis
\end{abstract}

\section{Introduction}
\fontsize{14}{14}\selectfont It's beyond doubt that the energy sector incorporates some of the world's most important topics and challenges in the 21st century. Humans have become utterly reliant on an ongoing stream of electricity, with constantly rising demand fueled by technology and innovation, such as the AI revolution. The need for energetic security and a sufficient energy supply puts topics like energy consumption and optimization at the center of attention. In addition, the effects of global warming caused by fossil fuels are perhaps humanity's biggest challenge. 

The energy and finance markets are closely related and interdependent. Since the 2008 global financial crisis, international energy markets have become more closely linked with financial markets, and energy prices have exhibited more financial characteristics \citep{energy_finance_background}. 
 A large number of studies investigate these new patterns and complex relationships, as well as issues such as the relationship between financing and investment decisions made by energy firms, carbon finance, energy derivatives, energy pricing mechanisms, and green finance. These studies have naturally evolved into a common research theme - energy finance, a subject of growing interest among academic researchers. \cite{energy_finance_background}\cite{frontiers_energy_finacne}.

Electricity market liberalization is a part of the broader process of energy market liberalization. It refers to changing the electricity market structure from state-owned and monopolized to an open market with individual players. Starting in the 1980s, countries around the world began changing their view and performing market reforms \citep{Implications_liberalization}. The old idea that electric power generation, transmission, and distribution represent  a  “natural  monopoly”  best  handled  centrally has  given  way  to  a  consensus  among  policy-makers, regulators, industry analysts, and economists that the generation and retailing elements of  the  power  supply  industry  would  be  more  efficiently  delivered  by  firms  operating in freely competitive energy markets \citep{trevino_liberalization}. The liberalization of the energy and electricity market has potential social, environmental, and financial effects. One of the main incentives for the reform is that a free market would deliver financial benefits to consumers. Therefore, understanding the relationships of the intervention with the financial markets, and more specifically, the causal impact on the electricity price, can be of high value for economists and policy-makers. 

Our research focuses on electricity market liberalization in the United States. In the 1990s, a group of states decided to open the electricity market as part of an adoption of a larger reform, while others decided to keep it monopolized and controlled. As a result, the rate of individual electricity producers jumped, altering the dynamics in the electricity market. Our experiment aims to determine whether this caused a significant effect on the electricity price, and if so, to quantify this effect. We propose Causal ML as a novel approach to conduct causality impact analysis of electricity market liberalization. Our approach has notable advantages over the Difference-in-Difference (DiD) method, which was used in previous analyses - it can estimate the individual treatment effect (ITE) in addition to the average treatment effect (ATE); it doesn't require linearity; the parallel trends assumption isn't required. We compare the performance of ML-based state-of-the-art models and utilize them to estimate the causal effect. We use placebo tests to confirm our assumptions about the causality structure in the data in order to verify the reliability of the results.

Contributions of this paper include:
\begin{itemize}
    \item We propose Causal ML as a novel approach for causality impact analysis of electricity market liberalization.
    \item We provide informative conclusions about and an estimate of the causal impact of electricity market liberalization on the price in the US. 
\end{itemize}

\section{Literature Review}
\label{sec:lit_review}

During the last 40 years, most developed countries have gone through reasonably comprehensive privatization, restructuring, and deregulation programs in sectors that were previously regulated monopolies and/or state-owned: airlines, telecommunications, mail and package delivery services, railroads, and other sectors. While these reforms have not always proceeded without controversy or led to the results predicted, the general trend of public policy has continued to support liberalization and to move forward with additional liberalization reforms in sectors that were once dominated by regulated legal monopolies\cite{lessons_elec_market_lib}.

The trend of deregulation did not skip the energy sector, and specifically, the electricity market. In the past, in almost every country, the electricity market evolved to be owned and run by a vertically integrated single monopoly firm. Such a firm would be either state-owned or private and subject to heavy regulation by the state. These monopolies controlled all components of the electricity supply chain - generation, transmission, distribution, and retail supply, making them the only available option for both residential and industrial consumers in a certain geographic area. 
Over time, the electricity sector came under pressure to undergo liberalization reforms in many countries. This pressure stemmed from high operating costs, high retail prices, and falling costs of production from new facilities driven by low prices for natural gas and the development of more efficient generating technologies. The idea that market reforms would result in greater operational efficiency, and the perceived reduction in costs and prices it would generate, became increasingly important to policymakers \cite{macro_effects_elec_lib}\cite{lessons_elec_market_lib}\cite{regul_reform_usa}. Another argument in favor of electricity market liberalization is the adoption of new technologies. A monopoly with no competitors naturally has less motivation and urgency to adopt new technologies that would benefit both the production side and the consumers. On the other hand, in an open market with free competition, the individual players must take action to protect their market share and keep their customer base. Such actions may include the adoption or creation of new technologies that would either give them an advantage over the competition or encourage them to align with the market's standards. For example, entities in the electricity industry have adopted flexible, change-oriented computer systems. The adoption of new technologies has impacted the economy, reducing operational and utility costs. The firms have also developed enhanced, tailored billing systems and real-time pricing of electric power. Some companies have adopted Utility Translation Systems software to obtain consumption data from massive power customers \cite{big_data_electric_utilities}. The adoption of and investment in renewable energy sources is an additional benefit of electricity market reform. Apart from the obvious positive environmental implications, renewables may also offer economic benefits by lowering electricity prices due to their reduced production costs.
Cheng et al. found that China's electricity market liberalization contributed to a significant increase in hydro power generation as its cost was lower than thermal power \cite{china_reform_renewable_production}. Solar and wind energy may require an additional measure, such as government subsidies to compensate for high initial investment or higher electricity prices, yet the adoption of such energies is also higher among individual players. As a result, since the late 1980s, many countries have pushed reforms in the electricity sector, using different methods. These included total privatization transfer of business entities or activities, from the public to private ownership, restructuring of some market regulations (business environment whereby interested parties, like investors, can easily enter or leave the market), and restructuring of electricity prices. Each of the scenarios had significant impacts on the adopting countries, hence becoming case studies for other nations, adapting them to their context \cite{deregulation_effects_overview}. 

From a customer perspective, the effect of electricity market liberalization on the price of electricity is one of the most important aspects, if not the most important one. Some research claims that while the benefit in price and service quality is quite obvious for large customers, the results are much less clear when it comes to small customers. Littlechild reviewed the experiences of retail competition, finding that large electricity customers are unequivocally better off because of liberalization, whereas when it comes to residential customers, some are better off and some are not \cite{evolution_competitive_markets}. 
As this paper focuses on residential prices, we are more interested in the effects on small customers. Several studies tried to examine which conditions are necessary for a reduction in residential price to occur. Such conditions include real-time pricing and time-of-use tariffs \cite{rethinking_realtime_elec_price}. Still, the literature in this regard remains quite inconclusive. 
In Spain, research concluded that liberalization of the electricity market caused an increase in electricity price and costs, though this increase is mainly attributed to a lack of adjustment strategy and an unforeseen effect in terms of the system's balance requirements \cite{spain_lib_effects}. Another study on panel data from 78 countries concluded that electricity prices are one of the driving forces for governments to adopt liberalization models. However, the development of liberalization models in the power sector does not necessarily reduce electricity prices. In fact, contrary to expectations, there was a tendency for the price to rise in every market modeled \cite{panel_data_elec_price_analysis}. Amenta et al. inspected the deregulation in the 27 EU member countries and found that competition is associated with lower prices for residential customers and higher efficiency. In addition, full liberalization has stronger effects than partial liberalization. \cite{eu_dereg_prices}. One conclusion that is evident and logical is that deregulation and liberalization of the electricity market must be made with consideration, rigorous planning, proper communication with consumers, and adequate transition measures to ensure customers can reap the benefits of the process and prevent negative impacts. 

Previous research was also conducted on the effects of market liberalization in the US. Su et al. concludes that customers in states that did go through market liberalization benefited from lower prices, yet that this benefit occurred only within the small customer sector and not in the industry and commercial sectors. Additionally, the benefit was only significant in the short term; in the long term, this effect vanished and became insignificant \cite{customer_benefit_usa}. One problem of this analysis and most literature focused on the US case is the fact that they fail to set the right timing when determining the intervention date. Commonly, the year of policy approval, or the policy approval year with an additional transitional period equal across all states, was used to set the intervention time. In reality, the time between the policy approval and its implementation was significantly different across states. This means that the timeline for private companies to enter the market and offer an alternative varied from state to state. In some states, the process took even ten years \cite{Retail_Electricity_Market_Restructuring}. A more comprehensive study, which included every state's restructuring timeline and accounted for price discounts during the transition period induced by the government, found evidence that electricity market liberalization reduced retail electricity prices in the short term, but no significant evidence for such reduction in the long term. \cite{Retail_Electricity_Market_Restructuring}. Yet this study, similar to the other studies conducted on this subject, uses the difference-in-differences method. Though a well-established and valid method, it forces linear and parametric assumptions as well as the parallel trend assumption, which may not be completely valid and adequate to describe the behavior of the electricity price. Therefore, investigating the case using new machine learning methods, which have no parametric assumptions and can understand non-linear relationships, can shed new light.

\subsection{Causality Analysis}
\label{subsec:frameworks}
The traditional statistical theory and its large set of tools and methods can explain associations between variables, identify patterns, and use past data to predict the future. The common factor for the above-mentioned functionalities is that they can learn the joint distribution and its parameters. Then, they utilize the distribution to perform the desired task. That is, they can answer research questions such as "How would Y behave given X?" However, as powerful as they are, the statistical methods are not able to tell us how a variable \textbf{changes} the distribution itself. Such analysis is not possible as there is nothing in the joint distribution of X and Y that can tell us how a change in X would affect the distribution of X and Y. The question "Is there a relationship of \textbf{causality} between X and Y?" is far more complex. In order to answer such questions, further assumptions must be made. These assumptions cannot be observed in the data alone and must be made using domain knowledge and expertise.
When the research is done in a supervised experiment setting, A/B testing or Randomized Control Trials (RCT) are efficient methods for estimating the causality effect. The randomness in RCT accounts for the confounding effect and produces unbiased estimates. The average treatment effect (ATE) can then be estimated as the difference in the dependent variable between the control and treatment groups \cite{rct_alzheimer}. Yet such experiments are often unfeasible due to ethical reasons and high costs. In addition, when the research concerns a given observational data, a different approach must be made. A common way to estimate the causal effect of an intervention is forecasting counterfactual time series outcomes based on similar units unaffected by the intervention at the same point in time. This artificial counterfactual trend is then used to estimate the causal effect of the intervention by differentiating the observed series affected by the intervention and the artificial counterfactual. These comparative case studies are ubiquitous in empirical research in the social sciences with different approaches \cite{deepprobcp}.

\section{Methodology}
\label{sec:methodology}
We aim to understand whether the electricity market liberalization in the United States \textit{caused} a change in the electricity price and, if so, to estimate the magnitude of this change. 
In this section, we discuss Rubin's framework, a prominent causality analysis framework that provides the principles and guidelines for the identification and estimation of a causal effect. We then narrow down the discussion to Synthetic Control, a popular method among Causal ML research that follows Rubin's framework. Then, we present the setup of our specific research question and elaborate on Causal ML models that are suitable for it.

Rubin's framework relies on the estimation of the causal effect by the potential outcome - the outcomes that would have happened had the treatment not occurred. These are widely known as the counterfactuals. Rubin's framework, also known as the Potential Outcome Framework (POF), extracts the causal effect by estimating the counterfactuals using the observed data and doesn't aim to build a full causal model. Rubin extended the concept of potential outcomes, initially proposed by Neyman for randomized experiments \cite{neyman_framework}, to observational studies, focusing on how statistical tools can be used to interpret a causal effect \cite{rubin_1974}\cite{rubin_1977}. The POF is comprised of three essential steps, involving the definition, identification, and estimation of causal effects. 
\begin{itemize}
\item \textbf{Definition:} Define causal estimands of interest, such as an average treatment effect, in terms of potential outcomes.
\item \textbf{Identification:} Identify observable statistical quantities that are equal to the defined causal effect, under particular assumptions.
\item \textbf{Estimation:} Estimate the identified observable statistical quantity using statistical models and related methodologies.
\end{itemize}

The POF assumes that each individual $i$  has two potential outcomes, $Y_i^1$ and $Y_i^0$, denoting the outcome when they receive treatment and when they do not, respectively. As a result, each individual’s observed outcomes can be written as:
\begin{align}
    \label{eq:obs_outcome}
    Y_i^{obs} =Z_iY_i^1 +\left(1 - Z_i\right)Y_i^0
\end{align}

Equation $1$ tells us that we observe the potential outcome under treatment $Y_i^1$ if an individual received the treatment $\left(Z_i = 1\right)$, and $Y_i^0$ in case the individual did not receive the treatment $\left(Z_i = 0\right)$. Thus, the outcome we observe is a function of an individual’s potential outcome and the treatment he receives. The main idea behind the POF is that the causal effect can be extracted by estimating the difference between the observed outcomes and the counterfactuals of the treated units. This can be formulated as $Y_i^1 - Y_i^0$.
The fact that for treated units we only observe $Y_i^1$ means that causality requires data that cannot be observed for its estimation. Indeed, without further assumptions, any causal estimand defined by potential outcomes cannot be identified by an estimable quantity. Below are the three assumptions required for the validity of the methodology.

\begin{itemize}
\item \textbf{Unconfoundedness:} Conditional on the covariates $X_i$, the treatment $Z_i$ is independent of the potential outcomes $\left(Y_i^1, Y_i^0\right)$, for all individuals.
\item \textbf{Positivity:} Conditional on the covariates $X_i$, every individual has some non-zero probability of receiving treatment or control. In other words, $0 < ps\left(X_i\right) < 1$, where ps is the probability, also known as the propensity score.
\item \textbf{Consistency:} Equation \ref{eq:obs_outcome}  holds, i.e., $Y_i^{obs} = Z_iY_i^1 + \left(1 - Z_i \right)Y_i^0$ for all subjects, regardless of how $Z_i = 1$ \text{or} $Z_i = 0$ is obtained. 
\end{itemize}
Unconfoundedness states that there is no information contained in subjects’ potential outcomes that informs how subjects receive treatment, beyond what information is contained in the covariates $X_i$. It is often a controversial assumption for observational studies, because it states that the covariates $X_i$ capture all variables needed to establish independence between $Z_i$ and the potential outcomes. Positivity states that every subject has some chance of receiving treatment or control. Consistency, also known as the Stable Unit Treatment Value Assumption (SUTVA) \cite{rubin_1986} states that if a
subject receives treatment $\left(Z_i = 1\right)$, then we will always see their treatment potential outcome $Y_i^1$. \cite{pot_outcome_framework}. \newline

The development of the potential outcome framework has given rise to significant research. Different ways to approach the potential outcome were further suggested, such as the Synthetic Control method developed in the early 2000s \cite{synth_control}.
The synthetic control (SC) method is perhaps the most widely used following Rubin's POF. It is suitable for the case of one or a few treated units, and therefore can be successfully implemented in policy intervention cases, where only a few treated units are available. The SC method addresses many of the DiD limitations \cite{DiD_model}, such as a parallel and linear trend between the control and treatment units. In addition, while DiD requires exogenous predictors to be specified in the model, the structure of SC allows the omission of such variables as long as their behavior remains the same across both control and treatment groups \cite{synthetic_control_examination}. The main idea behind SC is that a synthetic treated unit, which matches the treated unit's outcomes in the pre-treatment period, can be constructed by a weighted combination of control units. Consider a set of $N$ units indexed by $i = 1,...,N$. Let R denote the treated units with $1 \leq|R|<|N|$ and C the control (untreated) units with $|C| =|N|-|R|$. Let $T$ be the range of the time series index by $1,...,T_0,...T$ where $T_0$ represents the intervention time. Let $Y_{i\in{R},t}^{(1)}$ represent the observed outcomes in the presence of the intervention, and let $Y_{i,t}^{(0)}$ represent the observed outcomes in the absence of the intervention. Therefore, we first assume consistency - for the pre-treatment period, we observe $Y_{i,{t<t_0}}^{(0)}$ for both $i\in{R}$ and $i\in{C}$. On the other hand, in the post-treatment period, we observe $Y_{i\in{R},t\geq{T_0}}^{(1)}$ for \textit{all} treated units and $Y_{i\in{C},t\geq{T_0}}^{(0)}$ for \textit{all} control ones. The key assumption of SC is that the counterfactual of a treated unit $Y_{i\in{R},t\geq{T_0}}^{(0)}$ can be estimated as a function of a set of observed variables from a pool of covariates (control units and potentially additional external predictors). Therefore, the SC attempts to find the optimal combination of weights for the available covariates. Note that sometimes the "donor group" denoted by $D$ refers to the set of similar control units used to construct the synthetic treated unit, yet this is equal to referring to the full set of control units and assigning 0 to the weights of the unused units. Another key idea of this assumption is that the unobserved confounding impact on the treated unit can be balanced by a weighted combination of the unobserved confounding impact on the control units. 
The assumption can be formulated as following:\\
\textit{Assumption 1:}
\label{eq:synth_asmp_1}
\begin{equation}
    \mathrm{E}[Y_{i\in{R},t}^{(0)}] = \mathrm{E}\left[\sum_{j\in{C}}\left(\alpha_jY_{j,t}^{(0)} + \beta_jX_{j,t}\right)\right]
\end{equation}
Where $\alpha$ and $\beta$ are vectors of weights for the sets of control units $Y_{j\in{C}}$ and the treated unit external covariates $X_{i\in{R}}$, respectively. Note that this traditional linear structure can be extended to a non-parametric and non-linear structure of the form $\mathrm{E}\left[h\left(Y_{j\in{C},t}^{(0)},X_{j\in{C},t}\right)\right]$, where $h(\cdot)$ is a mapping function and $X_t$ is a set of external covariates containing also the treated unit's covariates. Another necessary assumption is that the relationship between a treated unit and its set of predicting variables modeled during the pre-intervention period remains the same after the intervention:\\
\textit{Assumption 2:}
\label{eq:synth_asmp_2a}
\begin{equation}
    Y_{i\in{R},t}^{(0)} = \lambda_0 + \gamma_t^{'}X_t +u_t \quad \forall t < T_0
\end{equation}

\label{eq:synth_asmp_2b}
\begin{equation}
    Y_{i\in{R},t}^{(1)} = \lambda_1 + \hat{\gamma}_t^{'}X_t +v_t \quad \forall t \geq T_0
\end{equation}

Where $X_t$ is the set of variables combining control units and external covariates, $\gamma$ is the vector of weights, and $u_t , v_t$ are the noise terms.

Equation 4 provides the necessary information to be able to model the counterfactual outcomes, similarly to equation \ref{eq:synth_asmp_2a}. Combining these assumptions allows us to estimate the unbiased counterfactual outcome: $$Y_{i\in{R},t}^{(0)}= \mathrm{E}\left[h\left(Y_{j\in{C},t}^{(0)},X_t\right)\right] \quad \forall t \geq{T_0}$$
And thus calculate the average treatment effect on the treated (ATT) at time t:
$$\pi_t = \frac{1}{n}\sum_{i=1}^{n}\left[Y_{i\in{R},t}^{(1)} - Y_{i\in{R},t}^{(0)}\right] = \frac{1}{n}\sum_{i=1}^{n}\left[Y_{i\in{R},t}^{(1)} - \mathrm{E}\left[h\left(Y_{j\in{C},t}^{(0)},X_t\right)\right]\right] \quad \forall t \geq T_0 $$

Note that confoundedness can lead to biased results and wrong conclusions about the causal effect of the intervention. Though by closely matching control units in the pre-treatment period, the SC method captures the effect of confounders, it assumes that no confounding effects influence the control and treatment differently over time. The existence of such time-varying confounders, or a failure to closely match control units in the pre-treatment period, results in the invalidity of the model. As in other cases, one must include such confounders using domain knowledge or assume unconfoundedness. As for exogenous predictors, failure to include such predictors with a dynamic nature (predictors whose behavior changes over time across control and treatment groups) can also lead to bias.

\textbf{\textit{The Global Approach}}

Recent research in the Causal ML field sought out to utilize the developments in Neural Networks (NN) and their ability to perform well in time series forecasting tasks. Grecov et al. proposed the global approach, a novel approach which is an extension of the traditional SC \cite{deepcpnet}. Traditionally, the SC method operates locally, which means that it optimizes the parameters for every treated unit separately. Additionally, the parameters learned optimize the weights given to the pool of control units and external covariates. In contrast, the global approach shares the parameters learned across all treated units, thereby increasing the shared information and enhancing the model's learning ability. Furthermore, the model is built as a time series forecasting predictor. Rather than predicting a treated unit's counterfactuals based on a weighted combination of control units and covariates, it uses the treated unit's pre-intervention data to predict the counterfactuals. 
Equation \ref{eq:synth_asmp_2a} tells us that for every time step before $T_0$ (the time of the intervention), we observe $Y_{i,t<T_0}^{(0)}$ for both treated and control units. Hence, the counterfactual estimation becomes the application of a forecast model to generate accurate predictions based on past time-series data: $$P_{i,t}^{(0)}= \mathrm h\left(Y_{i,{t<T_0}}^{(0)};\theta\right) \quad \forall t \geq{T_0}$$
where $h(\cdot)$ is a mapping function (the model), $Y_{i,{t<T_0}}^{(0)}$ a set of pre-intervention observed outcomes, and $\theta$ a set of learnable parameters.
The predictor is unbiased, that is, $\mathrm{E}\left( P_{i,t}^{(0)}\right)= \mathrm{E}\left( Y_{i,t}^{(0)}\right) \forall t \geq{T_0}$, thus allowing for an unbiased estimation of the potential outcomes of the treated units $ Y_{i\in{R},t}^{(0)} \forall t \geq{T_0}$. 
An additional aspect of this method is based on the assumption of consistency, specifically the null intervention effect on control units.  While for all treated units, we always observe the outcomes affected by the treatment in the post-treatment period  $Y_{i\in{R},{t >T_0}^{(1)}}$, for all control units, we observe outcomes unaffected by the treatment $Y_{i\in{C},{t >T_0}^{(0)}}$. This means that for control units, the predictions of our predictor should coincide with the observed outcomes. Thus, the global model approach uses the control units to asses the forecasting performance of the models by calculating the errors of the counterfactual prediction on the control only, which, under the null intervention assumption, should be similar to the observed outcome. Following that, we can verify the validity of the causal model using a placebo test. To construct the placebo test, we conduct a hypothesis testing of $H_0: \delta = 0$, where $\delta$ is the treatment effect. For the control units, we expect to fail to reject the null hypothesis. On the other hand, we expect a rejection of the null hypothesis for the treated units. In other words, the placebo test compares the errors of the predictions of the treated units (the treatment effect) to those of the control units. The distribution of the treatment units' errors should be substantially larger than the distribution of the control units' errors due to the treatment effect. If the placebo test is successfully passed, we can consider the TE estimation more reliable. 

We focus on the case of a single intervention that begins to affect all treated units in the same time step, which is defined as $T_0$. Additionally, we are interested in estimating the immediate short-term effect caused by the intervention.
Traditionally, the treatment effect of an intervention is regarded as a unified effect over all treated units. This is not always the case in real-life settings, where the magnitude of the effect often varies according to the distribution of the treated units. For example, a drop in oil prices may cause a sharper drop in the stock prices of small oil companies than it would in large ones, due to factors such as financial stability. In this paper, we explore such a case of heterogeneity in treatment effects and examine its effect on causal ML models' performance. 

\section{Experiments}
In this analysis, we employ several novel causal machine learning models and compare their performance. In the counterfactual prediction case, evaluating a model's ability to predict the counterfactuals is not feasible because they are unknown. For this reason, we first use a synthetic dataset, in which the treatment effect is known as it is artificially created. Such an approach allows for the precise evaluation of the models' ability to predict counterfactuals. Then, we proceed to implement the models on the real-world dataset.
While we cannot evaluate the models' predictions of the treated units (the counterfactuals), we expect our models to accurately predict the post-intervention outcomes of the control units. As these units are not affected by the intervention, the observed outcomes and the predictions of the models should align. 
Therefore, as an additional assessment step, in the real-world case, we compare the forecasting results of the control units only.

To evaluate the errors in the counterfactual estimation of the synthetic dataset and the prediction of control units in real-world data, we use the symmetric Mean Absolute Percentage Error (sMAPE) and the mean absolute scaled error (MASE). The two metrics are scale-invariant and are commonly used in time prediction evaluation. The metrics are defined as follows:
\begin{align}
    \mathrm{sMAPE} = \frac{2}{h}\sum_{t=n+1}^{n+h}\frac{\left|\hat{Y}_t-Y_t\right|}{\left|Y_t\right|+\left|\hat{Y}_t\right|}\hspace{0.25cm} \\
    \mathrm{MASE} = \frac{1}{h}\frac{\sum_{t=n+1}^{n+h}\left|\hat{Y}_t-Y_t\right|}{\frac{1}{n-s}\sum_{t=S+1}^{n}\left|Y_t-Y_{t-S}\right|}
\end{align} 
\subsection{Data}
\label{subsec:data}

\LARGE{{\textit{Synthetic Data}}}

\fontsize{14}{14}\selectfont The original idea of this synthetic dataset formula was created by Liu et al. \cite{pyraformer}. This dataset was chosen because it resembles the repeating seasonality structure of the real-world dataset in question, but with added complexity. It is constructed by a linear combination of three sine waves, a structure that is regarded as able to stimulate energy load data. To better fit our case, the generated dataset was modified from very long hourly data to a short daily dataset. Accordingly, the data-generating function was modified to incorporate daily (1), weekly (7), and monthly dependencies (30) and can be formulated as:
\begin{equation}
    f(t) = 100 +\beta_0 + \beta_1 sin(\frac{2\pi}{1}t) + \beta_2 sin(\frac{2\pi}{7}t) + \beta_3 sin(\frac{2\pi}{30}t),
    \label{eq:synth_function}
\end{equation}
Where $\beta_1$, $\beta_2$, and $\beta_3$ are uniformly sampled for each time point $t$ from [5,10]. $\beta_0$ is drawn from a Gaussian process that has a polynomially decaying covariance function $\Sigma_{t_1, t_2} = |t_1 - t_2|^{-1},$ where $\Sigma_{t_1} = \Sigma_{t_2} = 1$ and $t_1$ and $t_2$ are arbitrary time stamps. As the original function oscillates around zero, we added a constant of 100 to every value in the series. This step simply makes sure there are no negative values and does not affect any dependencies, temporal patterns, or any other property of the dataset. 

In addition to the stationary dataset created by the formula in equation \ref{eq:synth_function}, we also create datasets with a \textit{multiplicative} trend. The trend component we define is an exponential function of time such that $trend_t = trend\_rate^{t}$ for $t \in \{1, \ldots, \texttt{len(series)}\}$ and $y'_{t} = y_t \times trend_t$. The rate is set to $trend\_rate = 1.00005$.\\
\begin{figure}[h] 
  \centering
  \includegraphics[width=\textwidth]{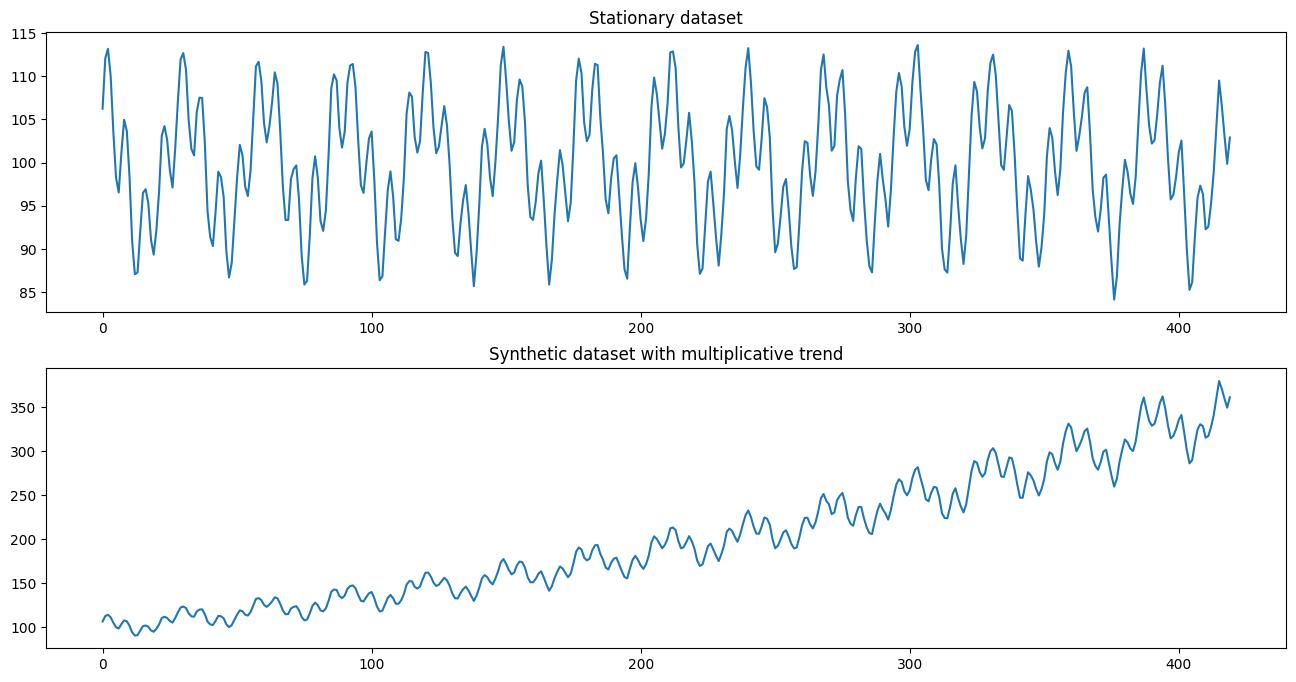} 
  \caption{Synthetic time series used in our experiments}
  \label{fig:synthetic_series}
\end{figure}
To simulate an intervention, we randomly split the units into control and treatment groups in a 7-3 ratio. We wish to create a scenario where the magnitude of the intervention's effect differs across the quantiles of the treated units' distribution, with higher quantiles affected more than lower ones. First, we calculate the standard deviation ($\sigma$) of the treated units' pre-intervention distribution. The intervention takes place in $T_0$, which is set as the length of the data minus 24 (to an intervention period that is similar to our real-world data). We then split the values of the treated units into 5 groups of two consecutive quantiles each and subtract a constant corresponding to a fraction of the calculated $\sigma$ from the values for all time steps after $T_0$. The constant changes based on the quantile group in which the value lies, in the following way: 

\begin{table}[h!]
    \centering
    \begin{tabular}{c|c}
    \toprule
       Quantile group  & Constant \\
    \midrule
        1st \& 2nd quantiles & 0.3$\sigma$ \\
        3rd \& 4th quantiles & 0.6$\sigma$ \\
        5th \& 6th quantiles & 0.9$\sigma$ \\
        7th \& 8th quantiles & 1.2$\sigma$ \\
        9th \& 10th quantiles & 1.5$\sigma$ \\
    \end{tabular}
    \label{tab:synth_constants}
\end{table}
\begin{figure}[h] 
  \centering
  \includegraphics[width=\textwidth]{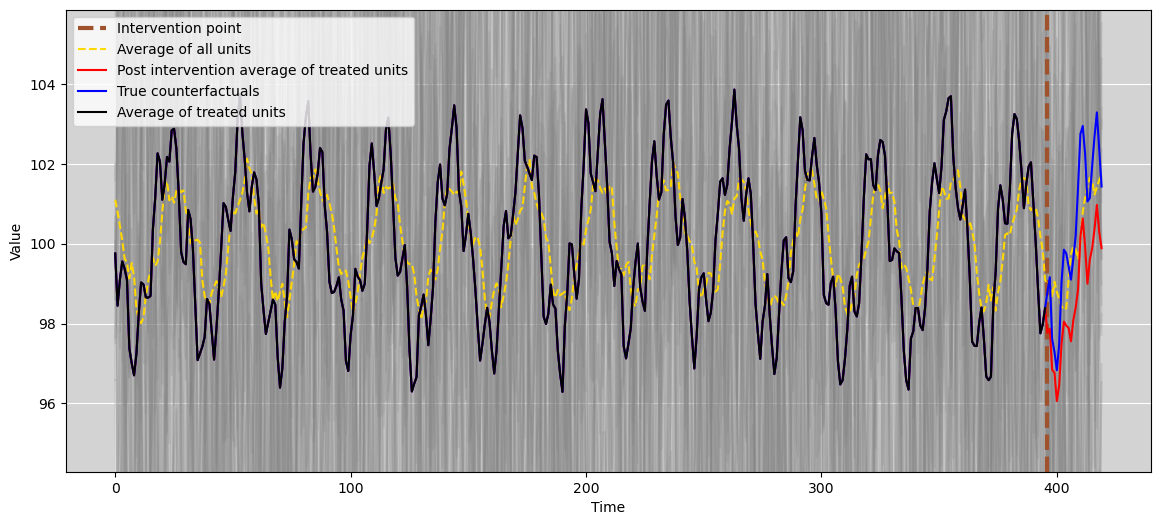} 
  \caption{Simulated dataset with intervention. The grey lines in the background are the time series. The dashed red line represents the average of all units, while the black line represents the average of the treated units. The blue and grey lines show the difference between the treated units post-intervention and the counterfactuals, respectively.}
  \label{fig:synth_intervention}
\end{figure}
We create combinations of small (50 time series) and large (300 time series) datasets with short (90 observations) and long (420 observations) lengths. Each combination is created both as stationary data and with an added trend. We then add the intervention to each simulated dataset.

\LARGE{\textit{Real-world data}}

\fontsize{14}{14}\selectfont We are interested in inspecting the causal impact of electricity market liberalization on electricity prices in the US. Starting from the late 90's, some states in the US introduced policies that altered the electricity market structure, which was dominated by a state-owned monopoly that governed the production, distribution, and supply of electricity. New regulations allowed individual players to join the market and sell electricity to customers. To empirically estimate the effect on the price, we gathered state-level data on electricity prices from the US Energy Information Administration (EIA) between 1990 and 2009 \cite{eia_dataset}. We extract the residential price data as we are interested in the effect on small customers. To obtain the state deregulation year, we checked data from the EIA website and cross-checked with available state website data. We found 17 states that passed electricity deregulation policies in the years 1996-2000. One observed error from previous studies was setting the intervention time as the year the policy was passed. In many cases, the policy came into effect only years after it was passed. This fact can be explained by the investment in infrastructure required to start operating, as well as the operational and legislative steps the state needed to perform after passing the laws. One could claim that the policy change caused a price difference even before it came into effect, as markets usually behave according to future assumptions. However, this claim can be ruled out in this case for two reasons: 1) the electricity market was dominated by a monopoly that had no reason to lower the price until an actual change occurred, and 2) the price data in all deregulated states does not change after the policy's approval. 
\begin{figure}[h] 
  \centering
  \includegraphics[width=\textwidth]{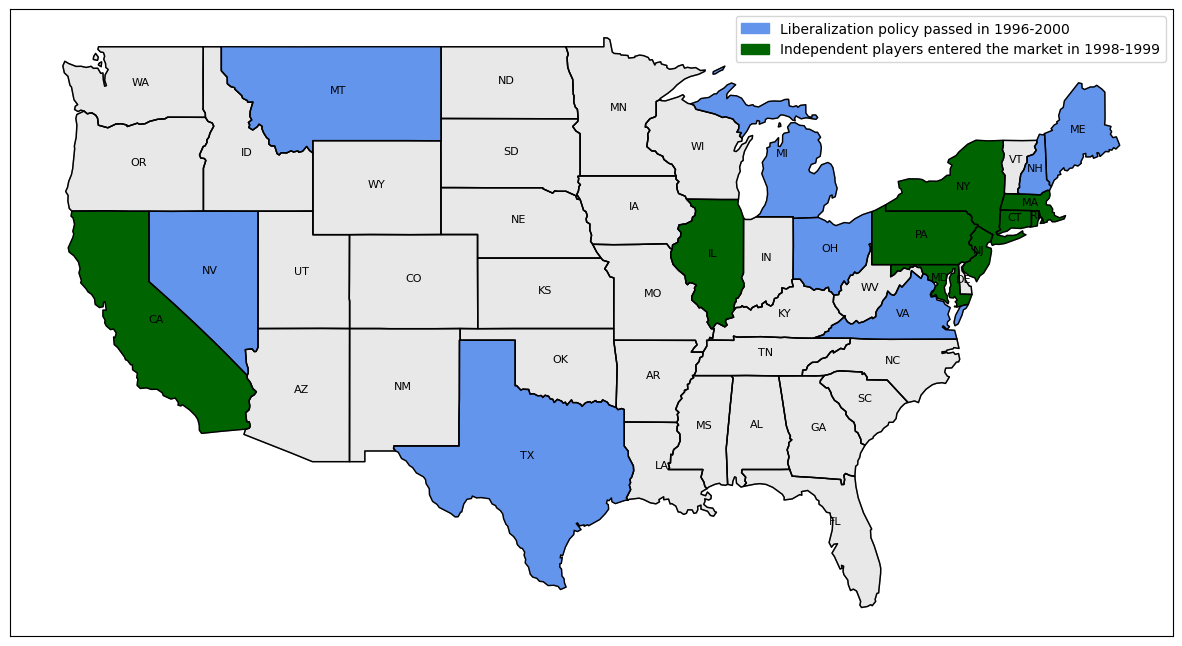} 
  \caption{17 states that passed deregulation policies between 1996 and 2000. Only 8 of them saw individual producers entering in 1998-1999.}
  \label{fig:usa_states}
\end{figure}
We used state-level annual data on electricity production by type of producer obtained from EIA \cite{eia_dataset} to calculate the share of individual producers of total production. Upon inspecting the price data, we can observe a visible change as the share of individual producers starts to go up. For the abovementioned reasons, we opt to use the year of the first spike in individual producers' share as the time of intervention. 

\begin{figure}[h] 
  \centering
  \includegraphics[width=\textwidth]{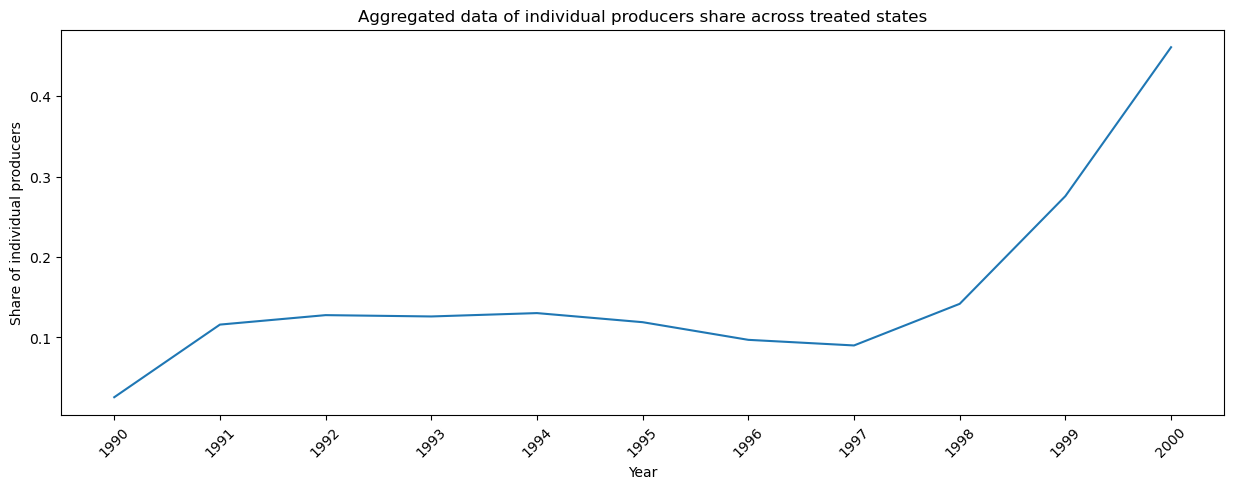} 
  \caption{The average of individual producers' share across treated states. A visible spike occurs starting from 1998}
  \label{fig:ind_prod_share}
\end{figure}
Our analysis deals with a single intervention that occurs at the same point in time across all treated units. For this reason, we group the liberalized states based on individual producers' market penetration year. Out of the 17 states, in 8 states, the share of individual producers jumped in 1998-1999: California, Connecticut, Illinois, Maine, Maryland, New Jersey, New York, Pennsylvania, and Rhode Island. The aggregated data of these states shows a significant increase in individual producers' market share during these years. We then set the time between 1990 and 1997 as the pre-intervention period and 1998-1999 as the post-intervention period. It's worth mentioning that the nature of our analysis is to use the pre-intervention period to estimate the short-term causal effect. Therefore, even though the share of individual producers keeps going up in 2000, this is out of the scope of this analysis. We therefore truncate all our data to the years 1990-1999. 
\begin{figure}[h] 
  \centering
  \includegraphics[width=\textwidth]{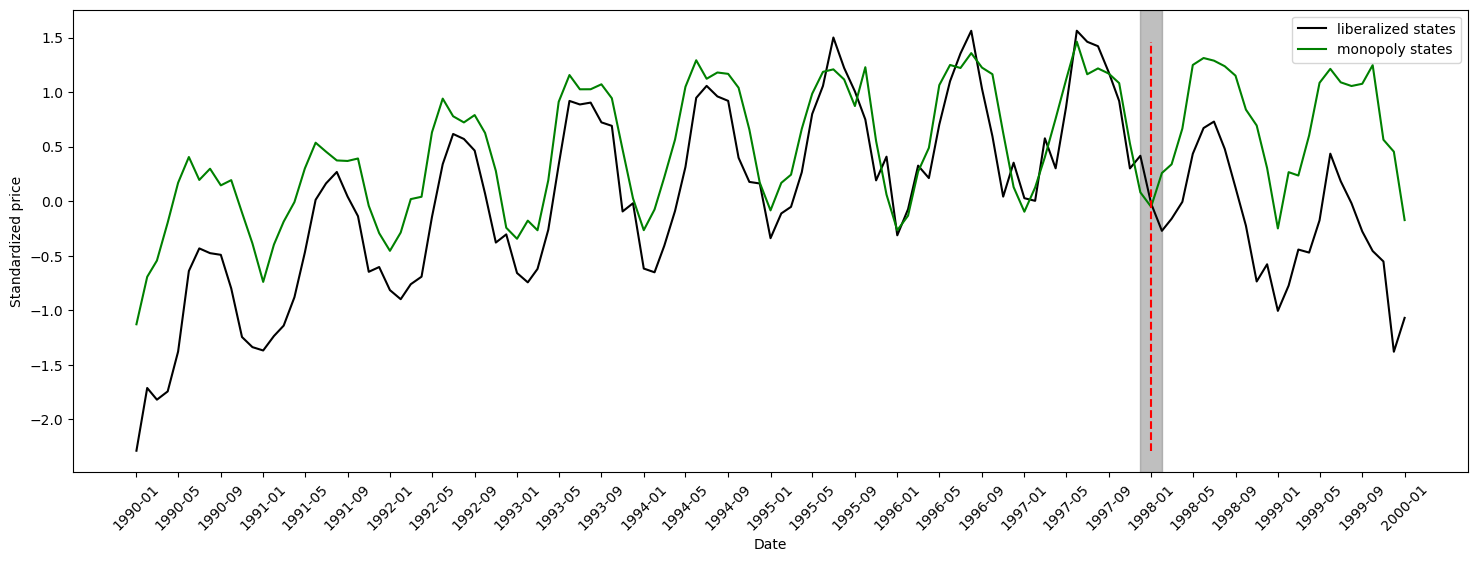} 
  \caption{Electricity price for monopolized and liberalized states. A visible drop in liberalized states occurs in 1998 when the share of individual producers begins to rise.}
  \label{fig:prices_graph}
\end{figure}
To account for additional factors that affect electricity prices, we add external covariates that represent the supply and demand. The idea is that including such representative covariates for both supply and demand should account for potential confounding effects on the electricity price and allow a proper estimation of the effect of market liberalization. To control for the demand side, we add the state-level aggregated average income obtained from the Bureau of Labor Statistics (BLS). For the supply side, following Su's work \cite{customer_benefit_usa}, we add the state-level gas price.

\subsection{Models}
We implement models from different branches to compare the performance of different methods, given that these methods are built in a way that is applicable to our problem. We use two synthetic control algorithms - \textit{DeepProbCP}, which works globally, and \textit{ASCM}, which is a local model. In addition, the \textit{TSMixer} is an MLP-based time series prediction framework that can be implemented in a synthetic control causal model. Lastly, the CausalArima is an econometric model that serves as a benchmark.
\begin{enumerate}

\item \textbf{DeepProbCP:} A global non-parametric DNN framework following the SC approach. It has a noticeable advantage over the traditional SC in the case where multiple treatment units exist. Instead of taking each treated unit separately and constructing the synthetic unit from the most similar control ones (the local approach), it learns the parameters of the predicting model by considering both treatment and control units together. By doing so, it utilizes all the available data as well as cross-series information. It uses the moving window strategy combined with the Seasonal Exogenous (SE) approach that extracts the seasonality component in the pre-training stage. The pre-intervention data of all units is used as training data, and the post-intervention period is used as the test data to be forecasted by the model.
The DeepProbCP has additional probabilistic forecasting capabilities. Unlike standard point estimation, it can predict different quantiles of the counterfactual distribution, hence providing a wider image of the causal effect. The model is trained with an LSTM layer and uses Continuous Coin Betting (COCOB) optimizer to minimize the Quantile Loss (QL) per moving window. It uses the Continuous Ranked Probability Score (CRPS) as its optimization metric. The model is implemented using the DeepProbNet \cite{deepcpnet} code provided by the authors.
\item \textbf{TSMixer:} The TSMixer is an MLP-based framework that achieved state-of-the-art results in multivariate time series forecasting problems. It received a lot of attention thanks to the fact that it managed to outperform attention-based Transformers models while also having a much lower complexity that renders its implementation more accessible. The model uses two MLPs in parallel - the first works along the time axis to find temporal patterns and is shared across the features, while the second operates along the feature axis and is shared across the time steps. This structure allows the model to successfully extract both patterns and information created over time, as well as complex relationships between features. We utilize the prediction abilities of the framework as a part of our causal model to predict the counterfactuals. Similar to the DeepProbCP framework, the model uses the moving window strategy, and the parameters are learned globally across both control and treated units in the pre-treatment period. The TSMixer was developed by Google Research \cite{tsmixer} and the model is implemented using the publicly available code. 

\item \textbf{ASCM:} The Augmented Synthetic Control Method (ASCM) is an extension of the original SC method proposed for the (common) cases where a good fit of the treated unit amongst the control ones is not achievable \cite{augmented_synh_control}. It first works like the standard SC model and creates the synthetic unit from a pool of control units. Then, it estimates the bias created by imperfect fit and debiases the original estimation. The main idea is to add a Ridge regression model, which, unlike the standard SC model, allows for negative weights for some control units and uses extrapolation to improve the pre-treatment fit. When the estimated bias is small, the ASCM and the standard SCM estimates will be similar. When the bias is large, however, the ASCM will rely more on extrapolation. This method is local and can deal with one treated unit at a time. Though it does not use more advanced techniques such as neural networks, it is nevertheless a potential novel alternative. The ASCM framework is implemented using the \textit{augsynth} R package. 
\item \textbf{Causal ARIMA:} We use the C-ARIMA as a benchmark model that uses econometric methods for time series prediction \cite{c-arima}. This framework bridges the gap between causal inference under the potential outcome framework and intervention analysis using ARIMA models. It makes use of ARIMA models to estimate the counterfactual prediction of a unit in an observational time series setting. We implement the model using the CausalArima R package.

\end{enumerate}

\section{Results}
\label{sec:results}

Table \ref{tab:synth_metric_results} summarizes the error metrics of the synthetic experiments. The errors refer to both the treated units and the control ones. The errors on the treated units are calculated using the true counterfactuals taken from the simulation data. The best results are bolded. The global DeepProbCP model appears to perform best in all cases of stationary data and most cases of data with a trend. The performance of the TSMixer is somewhat surprising, as this model proved to outperform many other NN-based models. This behavior could be because this model was built for longer prediction horizon cases of long time series data. It is evident from the results that the TSMixer's performance worsens as the length of the dataset gets shorter. In fact, for the short time series, its performance is the worst across all models. On the other hand, in the case of the longer time series, the TSMixer's performance gets significantly better. It is consistently better than the ASCM and C-Arima models, and its errors are almost as low as DeepProbCP's errors. It is likely that if the length of the time series were further extended, the structure of the TSMixer would allow it to outperform the other models. The ASCM method that utilizes Ridge regression to unbias the selection bias of the synthetic control unit generally presents the worst performance. The other local model, the C-Arima, is a better choice if a local model is desired. 

\definecolor{lightgray}{gray}{0.9}
\definecolor{lightgreen}{rgb}{0.6, 0.8, 0.4}

\begin{table}[ht]
\centering
\renewcommand{\arraystretch}{2} 
\begin{longtable}{@{}|c|c|c|cc|cc|cc|cc|@{}}
\caption{Forecasting error results on synthetic datasets}
\label{tab:synth_metric_results} \\

\toprule
\multicolumn{3}{|c|}{Model} & \multicolumn{2}{c|}{TSMixer} & \multicolumn{2}{c|}{DeepProbCP} & \multicolumn{2}{c|}{ASCM} & \multicolumn{2}{c|}{CArima}  \\
\midrule
\endfirsthead
\multicolumn{1}{|c|}{DGP} & \multicolumn{1}{c|}{Series} & \multicolumn{1}{c|}{Length}  & \multicolumn{1}{c}{sMAPE} & \multicolumn{1}{c|}{MASE} & \multicolumn{1}{c}{sMAPE} & \multicolumn{1}{c|}{MASE} & \multicolumn{1}{c}{sMAPE} & \multicolumn{1}{c|}{MASE} & \multicolumn{1}{c}{sMAPE} & \multicolumn{1}{c|}{MASE} \\
\midrule
\multirow{4}{*}{\rotatebox[origin=c]{90}{$\mathrm{Stationary}$}} & \multirow{2}{*}{50} & {90} & {0.085} & {1.214} & {\textbf{0.016}} & {\textbf{0.231}} & {0.063} & {0.905} & {0.05} & {0.737}   \\
\cline{3-11}
& & {420} & {0.025} & {0.352} & {\textbf{0.018}} & {\textbf{0.243}} & {0.066} & {0.902} & {0.037} & {0.504}   \\
\cline{2-11}
&\multirow{2}{*}{300} & {90} & {0.092} & {1.257} & {\textbf{0.018}} & {\textbf{0.24}} & {0.067} & {0.9} & {0.05} & {0.672}  \\
& & {420} & {0.023} & {0.322} & {\textbf{0.017}} & {\textbf{0.239}} & {0.065} & {0.896} & {0.038} & {0.522}  \\
\midrule

\multirow{4}{*}{\rotatebox[origin=c]{90}{$\mathrm{Trend}$}} & \multirow{2}{*}{50} & {90} & {0.103} & {1.138} & {\textbf{0.051}} & {\textbf{0.562}} & {0.067} & {0.758} & {0.061} & {0.682}   \\
\cline{3-11}
& & {420} & {\textbf{0.025}} & {\textbf{0.403}} & {0.027} & {0.438} & {0.071} & {1.175} & {0.027} & {0.451}   \\
\cline{2-11}
& \multirow{2}{*}{300} & {90} & {0.102} & {1.134} & {\textbf{0.047}} & {\textbf{0.528}} & {0.068} & {0.767} & {\textbf{0.047}} & {\textbf{0.526}} \\
& & {420} & {0.029} & {0.478} & {\textbf{0.026}} & {\textbf{0.429}} & {0.075} & {1.23} & {0.028} & {0.459}  \\
\bottomrule
\end{longtable}
\end{table}

Table \ref{tab:synth_att_results} presents the Average Treatment Effect on Treated (ATT). The estimated ATT of each model is calculated by the difference between the model treatment units' predictions and the true counterfactuals. The sMAPE is calculated only on the treated units' counterfactuals. In the short series, the estimated ATT of the DeepProbCP is again most similar to the artificially created  ATT. In the cases of the long series, it is less clear which model performs best. Methodologically, we should use the model that produces the lowest errors on the ATT in the cases most similar to the real-world dataset. In this sense, the DeepProbCP model is the most adequate one, because our real-world dataset is rather short and has 50 series.

\centering
\renewcommand{\arraystretch}{1.7} 
\begin{longtable}{c|c|c|c|ccccc|}
\caption{Average Treatment Effect on Treated (ATT) and the corresponding sMAPE results} 
\label{tab:synth_att_results} \\

\toprule
\multicolumn{4}{|c|}{Model} &  \multicolumn{1}{c|}{Real} & \multicolumn{1}{c|}{TSMixer} & \multicolumn{1}{c|}{DeepProbCP} & \multicolumn{1}{c|}{ASCM} & \multicolumn{1}{c|}{CArima}  \\
\midrule

\multicolumn{1}{c|}{} & \multicolumn{1}{|c|}{DGP} & \multicolumn{1}{c|}{Series} & \multicolumn{1}{c|}{Length} & \multicolumn{5}{c|}{} \\  

\midrule
\endfirsthead
\pagebreak
\multicolumn{9}{c}{Table \thetable\ continued from previous page} \\
\toprule
\multicolumn{4}{|c|}{Model} &  \multicolumn{1}{c|}{Real} & \multicolumn{1}{c|}{TSMixer} & \multicolumn{1}{c|}{DeepProbCP} & \multicolumn{1}{c|}{ASCM} & \multicolumn{1}{c|}{CArima}  \\
\midrule

\multicolumn{1}{c|}{} & \multicolumn{1}{|c|}{DGP} & \multicolumn{1}{c|}{Series} & \multicolumn{1}{c|}{Length} & \multicolumn{5}{c|}{} \\  
\endhead
\multirow{8}{*}{%
    \begin{tikzpicture}[overlay, remember picture]
        \fill[lightgray] (-1ex,-14.5ex) rectangle (2.5ex,16.0ex);
    \end{tikzpicture}%
    
\rotatebox[origin=c]{90}{$\mathrm{\textbf{ATT}}$}} & \multirow{4}{*}{\rotatebox[origin=c]{90}{$\mathrm{Stationary}$}} & \multirow{2}{*}{50} & {90} & {-6.72} & {-8.13} & {-6.85} & {-5.27} & {-8.34} \\

& & & {420} & {-7.11} & {-7.97} & {-7.73} & {-5.88} & {-7.26}  \\
\cline{3-9}
& &  \multirow{2}{*}{300} & {90} & {-7.06} & {-6.34} & {-7.04} & {-4.71} & {-7.34} \\
& & & {420} & {-6.91} & {-6.86} & {-7.15} & {-4.68} & {-6.99} \\
\cline{2-9}
&  \multirow{4}{*}{\rotatebox[origin=c]{90}{$\mathrm{Trend}$}} & \multirow{2}{*}{50} & {90} & {-13.76} & {-5.03} & {-7.59} & {-10.06} & {-9.56} \\

& & & {420} & {-24.06} & {-24.11} & {-21.27} & {-17.25} & {-22.38}  \\
\cline{3-9}
& &  \multirow{2}{*}{300} & {90} & {-13.45} & {-9.59} & {-8.26} & {-9.43} & {-12.56} \\
& & & {420} & {-24.07} & {-23.25} & {-21.28} & {-16.87} & {-22.68} \\
\midrule
\pagebreak
\midrule
\multirow{8}{*}{%
        \begin{tikzpicture}[overlay, remember picture]
        \fill[lightgreen] (-1ex,-14.5ex) rectangle (2.5ex,16.0ex);
    \end{tikzpicture}%

    \rotatebox[origin=c]{90}{$\mathrm{\textbf{sMAPE}}$}} & \multirow{4}{*}{\rotatebox[origin=c]{90}{$\mathrm{Stationary}$}} & \multirow{2}{*}{50} & {90} & {} & {0.086} & {\textbf{0.018}} & {0.064} & {0.053} \\

& & & {420} & {} & {0.029} & {\textbf{0.018}} & {0.067} & {0.037}  \\
\cline{3-9}
& &  \multirow{2}{*}{300} & {90} & {} & {0.093} & {\textbf{0.019}} & {0.067} & {0.05} \\
& & & {420} & {} & {\textbf{0.023}} & {0.017} & {0.065} & {0.039} \\
\cline{2-9}
&  \multirow{4}{*}{\rotatebox[origin=c]{90}{$\mathrm{Trend}$}} & \multirow{2}{*}{50} & {90} & {} & {0.109} & {\textbf{0.053}} & {0.068} & {0.065} \\

& & & {420} & {} & {0.023} & {0.025}& {0.069} & {\textbf{0.022}}  \\
\cline{3-9}
& &  \multirow{2}{*}{300} & {90} & {} & {0.102} & {\textbf{0.046}} & {0.068} & {\textbf{0.046}} \\
& & & {420} & {} & {0.028} & {0.026} & {0.075} & {\textbf{0.025}} \\
\bottomrule
\end{longtable}

\raggedright
\textit{\LARGE{Real-world data}}

Table \ref{tab:elec_price_errors} shows the prediction errors on the electricity price of the different models. The p-value represents the results of the Wilcoxon test for the difference in means of the control and treatment groups (the placebo test). A p-value smaller than 0.05 means the model passed the test. As we can see, all models successfully pass the placebo test; hence, we can deem their treatment effect estimation reliable. 
The errors (sMape and MASE) are calculated on the control units only, based on the null intervention on the control units assumption. Because the control units should not be affected by the intervention, we expect the models' predictions to be similar to the observed values of these units. Similar to the results from the synthetic datasets, the DeepProbCP model performs best with the lowest errors, both in terms of sMAPE and MASE. These results align with the previous experiments, as our real-world dataset is rather short (120 time steps) and contains 50 series. In a similar synthetic setting, the DeepProbCP model presented clear superiority over the other models. These results lead to some interesting findings - despite the demonstrated superiority of the TSMixer model in many time series forecasting settings, it is not necessarily fit for a causal model with intervention effect estimation. Such a model often deals with shorter time-series lengths and prediction horizons, making it harder for the TSMixer, which was built for long-range time-series predictions, to generate accurate forecasts. On the contrary, the DeepProbCP was tailored for such cases. By utilizing its seasonal exogenous methodology, with an LSTM layer and CRPS optimization, it is a novel method for the estimation of the short-term effect of an intervention. Generally speaking, the ASCM method presents the worst results. If a local method is preferred for a certain situation, we recommend using the C-Arima model as it outperforms the ASCM method in all cases.

\renewcommand{\arraystretch}{2} 
\begin{longtable}{@{}|c|cc|cc|cc|cc|@{}}
\caption{Forecasting error results on electricity price dataset (control units only)}
\label{tab:elec_price_errors} \\
\toprule
\multicolumn{1}{|c|}{Model} & \multicolumn{2}{c|}{TSMixer} & \multicolumn{2}{c|}{DeepProbCP} & \multicolumn{2}{c|}{ASCM} & \multicolumn{2}{c|}{CArima}  \\
\midrule
\endfirsthead
\multicolumn{1}{|c|}{P-Value} & \multicolumn{2}{c|}{0.003} & \multicolumn{2}{c|}{0.001} & \multicolumn{2}{c|}{0.000} & \multicolumn{2}{c|}{0.001} \\
\midrule
\multicolumn{1}{|c|}{sMAPE} & \multicolumn{2}{c|}{0.052} & \multicolumn{2}{c|}{\textbf{0.035}} & \multicolumn{2}{c|}{0.061} & \multicolumn{2}{c|}{0.041}  \\
\midrule
\multicolumn{1}{|c|}{MASE} & \multicolumn{2}{c|}{2.191} & \multicolumn{2}{c|}{\textbf{1.341}} & \multicolumn{2}{c|}{2.719} & \multicolumn{2}{c|}{1.852}  \\
\midrule
\bottomrule
\end{longtable}

\begin{figure}[h] 
  \begin{centering}
  \includegraphics[width=\textwidth]{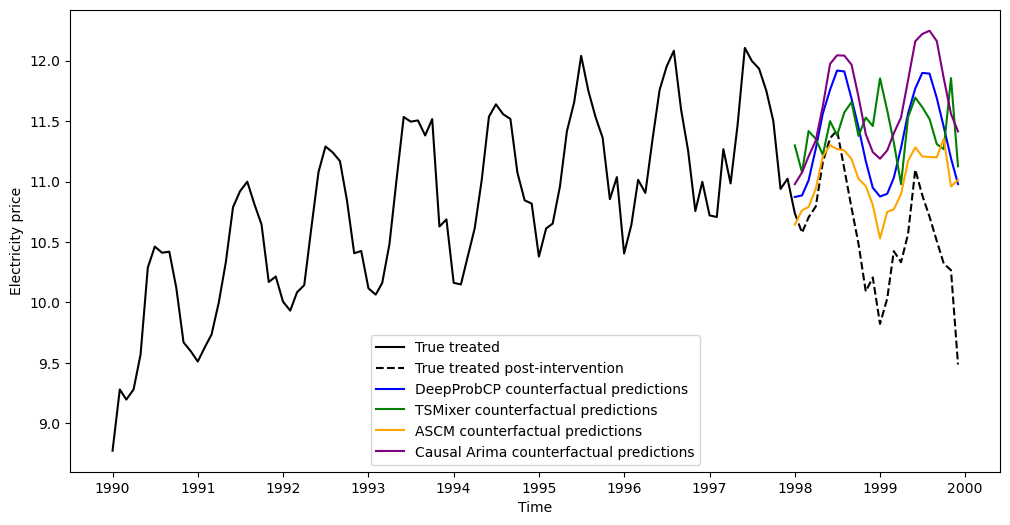} 
  \caption{Counterfactual predictions of all models.}
  \label{fig:all_model_predictions}
  \end{centering}
\end{figure}

In table \ref{tab:elec_price_att_results}, we can see the average treatment effect on treated estimation of the different models. Due to its lowest forecasting errors, we choose the DeepProbCP estimation as our estimation for this analysis. Since all models passed the placebo test and estimated a negative treatment effect, we can have more confidence in the results, which point out a clear price reduction in the post-intervention period for the treated units. For 2 years following the liberalization of the electricity market and the entrance of individual electricity providers to the market, the price decreased on average by 0.795 ¢/kWh which amounts to a 7\% price reduction compared to the average price in the year before the intervention. Though not a huge change, this finding does prove that open competition and individual electricity players contribute to more competitive electricity prices and benefit small residential customers.

\begin{centering}
\begin{longtable}{c|cc|cc|cc|cc}
\caption{ATT estimation of electricity market liberalization on electricity price} 
\label{tab:elec_price_att_results} \\
\multicolumn{1}{|c|}{Model} & \multicolumn{2}{c|}{TSMixer} & \multicolumn{2}{c|}{DeepProbCP} & \multicolumn{2}{c|}{ASCM} & \multicolumn{2}{c|}{CArima}  \\
\endfirsthead
\toprule
\midrule
\multicolumn{1}{|c|}{} & \multicolumn{2}{c|}{-0.859 $\mbox{\textcent}/kWh$} & \multicolumn{2}{c|}{-0.795$\mbox{\textcent}/kWh$} & \multicolumn{2}{c|}{-0.441$\mbox{\textcent}/kWh$} & \multicolumn{2}{c|}{-1.064$\mbox{\textcent}/kWh$}  \\
\bottomrule
\end{longtable}
\end{centering}

\begin{figure}[h] 
  \begin{centering}
  \includegraphics[width=\textwidth]{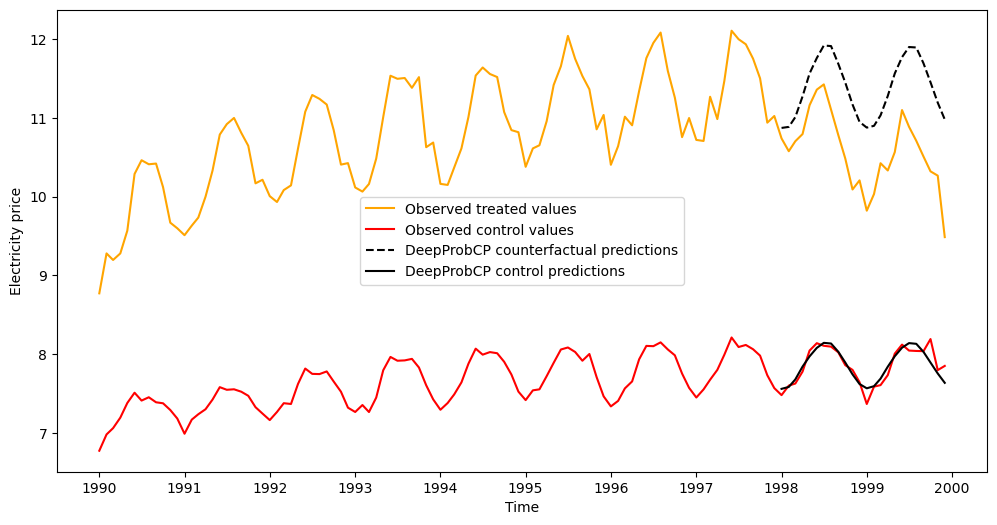} 
  \caption{DeepProbCP forecasts of treatment and control units. The treatment units' forecasts represent the counterfactuals for the ATT estimation}
  \label{fig:deepprobcp_elec_price}
  \end{centering}
\end{figure}

\section{Conclusion and Discussion}
In this paper, we aim to estimate the immediate short-term effect of electricity market liberalization on electricity prices for residential customers in the US. Through this analysis, we also discuss machine learning-based methods for causality analysis and highlight novel models. 
We choose frameworks that are adequate for the estimation of the short-term effect of a single intervention on multiple treated units and compare their performance. We first create multiple synthetic scenarios to assess the models' abilities to fit various data lengths and sizes, and continue to estimate the electricity market liberalization effect on the electricity price. Our work leads us to reach two conclusions. First, we find the LSTM-based DeepProbCP framework to be the most suitable for our research question, due to its consistently most accurate predictions of the counterfactuals. This framework also contains probabilistic quantile predictions that weren't explored in this paper and could reap further benefits and insights. The DeepProbCP has an established, straightforward causal model structure, but its implementation could be potentially improved with more advanced predictive models. Though the TSMixer shows inferior performance in our work, it is evident that as the length of the time series increases, its performance improves. Therefore, it is possible that for longer time series, the TSMixer architecture would outperform the rest of the models. Investigating the combination of the causal model framework of DeepProbCP with the forecasting MLP network of the TSMixer on longer time series could show superior results. The ASCM model, despite its bias-balancing structure,  did not demonstrate good prediction performance. For a local model, we propose the CausalArima model, which demonstrated consistently better performance. Additionally, we find that the liberalization of the electricity market and the entrance of individual producers led to a price decrease of 7\% on average during the first two years following the entrance of these individual players. Our results shed further light on the effects of liberalization policy and can help policymakers in the future. To further continue this direction of research, one can explore the long-term effect or investigate the potential relationships of the electricity price with price policies introduced by the liberalized states after the intervention.

\section{Acknowledgement}
This paper is supported through the project the projects "AI for Energy Finance (AI4EFin)", CF162/15.11.2022, contract number CN760048/23.05.2023 and "IDA Institute of Digital Assets", CF166/15.11.2022, contract number CN760046/ 23.05.2023, financed under the Romania’s National Recovery and Resilience Plan, Apel nr. PNRR-III-C9-2022-I8; and the Marie Skłodowska-Curie Actions under the European Union's Horizon Europe research and innovation program for the Industrial Doctoral Network on Digital Finance, acronym DIGITAL, Project No. 101119635.

\newpage
\bibliography{references}

\end{document}